%% file: acl_latex.tex
\pdfoutput=1

\documentclass[11pt]{article}

\usepackage[final]{acl}

\usepackage[utf8]{inputenc} 
\usepackage[T1]{fontenc}    
\usepackage{hyperref}       
\usepackage{url}            
\usepackage{booktabs}       
\usepackage{amsfonts}       
\usepackage{nicefrac}       
\usepackage{microtype}      
\usepackage{lipsum}         
\usepackage{graphicx}
\usepackage{xcolor}
\usepackage{soul}
\usepackage{doi}
\usepackage{listings}
\usepackage{multicol}
\usepackage{array}
\usepackage{tablefootnote}
\usepackage{multirow}
\usepackage{xspace}
\usepackage{xcolor}
\usepackage{svg}
\usepackage{amsmath}
\usepackage{cleveref}       
\usepackage{adjustbox}
\usepackage{colortbl}
\usepackage{float}
\usepackage{placeins}
\usepackage{titlesec}
\input{math_commands.tex}

\usepackage{amsmath}
\usepackage{pgfplots}
\pgfplotsset{compat=newest}
\usepackage{enumitem}

\newcommand{\tool}{\textsc{RepoHyper}\xspace}

\definecolor{XF}{RGB}{207,226,243}
\definecolor{IF}{RGB}{255,242,204}
\definecolor{NL}{RGB}{217,234,221}
\usepackage{times}
\usepackage{latexsym}

\usepackage[T1]{fontenc}

\usepackage[utf8]{inputenc}

\usepackage{microtype}

\usepackage{inconsolata}

\title{\textbf{\texttt{\textcolor{blue}{RepoHYPER}}}: : Search-Expand-Refine on  Semantic Graphs for \\ Repository-Level Code Completion}

 \author{Huy N. Phan$^{\spadesuit}$ \quad Hoang N. Phan$^{\clubsuit}$ \quad Tien N. Nguyen$^{\diamondsuit}$\quad Nghi D. Q. Bui$^{\heartsuit}$ \\
         $^{\spadesuit}$FPT Software AI Center, Viet Nam \\
         $^{\clubsuit}$Department of Computer Science, Nanyang Technological University, Singapore \\
         $^{\diamondsuit}$Computer Science Department, The University of Texas at Dallas, USA \\
         $^{\heartsuit}$Fulbright University, Viet Nam \\
         \texttt{huypn16@fpt.com, C210055@ntu.edu.sg, tien.n.nguyen@utdallas.edu, nghi.bui@fulbright.edu.vn}}

\begin{document}

\maketitle
\begin{abstract}

Code Large Language Models (CodeLLMs) have demonstrated impressive proficiency in code completion tasks. However, they often fall short of fully understanding the extensive context of a project repository, such as the intricacies of relevant files and class hierarchies, which can result in less precise completions. To overcome these limitations, we present \tool, a multifaceted framework designed to address the complex challenges associated with repository-level code completion. Central to \tool is the {\em Repo-level Semantic Graph} (RSG), a novel semantic graph structure that encapsulates the vast context of code repositories. Furthermore, \tool leverages \textit{Expand and Refine} retrieval method, including a graph expansion and a link prediction algorithm applied to the RSG, enabling the effective retrieval and prioritization of relevant code snippets. Our evaluations show that \tool markedly outperforms existing techniques in repository-level code completion, showcasing enhanced accuracy across various datasets when compared to several strong baselines. Our implementation is published at. \footnote{https://anonymous.4open.science/r/RepoHyper-3836/README.md}
\end{abstract}

\input{latex/introduction}

\input{latex/relatedwork}

\input{latex/methodology}

\input{latex/experiment-setup}

\input{latex/results-analysis}

\input{latex/conclusion}

\newpage

\input{latex/limitations}

\bibliography{latex/custom}

\appendix

\section{Appendix}
\label{sec:appendix}

\subsection{Building RSG}
In this section, we present details on how to build nodes and relations of Repo-level Semantic Graph.
Firstly, we use Tree-sitter \footnote{https://tree-sitter.github.io/tree-sitter/} to parse functions, methods, classes out of Python files. After parsing these entities, we removed these entities from each file so the remaining codes in the file is not function or class. This ensures import statements and other statements (like main file) will be remained. In order to create \textit{Import Relations} between script nodes and the imported nodes (functions, classes), we use built-in Abstract Syntax Tree (AST) module of Python to parse import statements, then identify which module in parsed entities is imported into the script node.
For \textit{Invoke Relations}, we use PyCG \footnote{https://github.com/vitsalis/PyCG} to generate the call graph of the repository. Since, PyCG only works for Python3 repositories, and RepoBench contains a lot of Python2 repositories, we use Automated Python 2 to 3 code translation tool 2to3 \footnote{https://docs.python.org/3/library/2to3.html} to translate these repositories into Python3, then apply PyCG to produce call graph. \textit{Ownership} and \textit{Encapsulate} relationships are straightforward to generate since the Tree-sitter allows us to identify which method belongs to which class exactly and we also parse functions, classes from each file then we also know which function, class belongs to which file exactly. For \textit{Class Hierarchy Relations}, since Python is a language that needs to specify parent class in the implementation of the inherited class, we can use Tree-sitter to parse this parent class in declaration fields of the inherited class, then we can produce a class hierarchy edge between parent and child class.
\label{sec:appendix-rsg}

\subsection{Gold Snippet Definition}
In the training set of RepoBench-R, for each sample, every snippet parsed from import statements is treated as a potential candidate for next-line prediction, for example, we have definitions of two following imported functions: add and minus as the snippets and the gold snippet is the optimal context for prediction, here which is "add" from src import add, minus nextline: add(1,2).

\subsection{Pattern Search}
\label{appendix:ps}
In Repository Semantic Graph (RSG), there are three types of nodes: function (method), class, and script. The edges between these nodes represent five different relationships: import (\texttt{1}), invoke (\texttt{2}), ownership (\texttt{3}), encapsulate (\texttt{4}), and hierarchy (\texttt{5}). A path type is defined by the sequence of edge types along that path in the RSG. For example, let's consider a node $V_j$ identified by the $k$-nearest neighbor (kNN) search, and a target node $V_{\text{target}}$ representing the ``gold snippet" (e.g., a method called by $V_j$). There can be multiple paths from $V_j$ to $V_{\text{target}}$. One such path could be: $V_j$ (method) $\rightarrow$ (owned by) $V_1$ (class) $\rightarrow$ (encapsulated in) $V_2$ (script) $\rightarrow$ (import) $V_{\text{target}}$ This path has a type of (ownership, encapsulate, import), represented by the edge types (3, 4, 1). However, since there are multiple paths from $V_j$ to $V_{\text{target}}$, including all the nodes produced by these paths might introduce irrelevant contexts into the subgraph. For example, $V_j$ can reach $V_{\text{target}}$ through a more useful path like: $V_j$ (method) $\rightarrow$ (owned by) $V_1$ $\rightarrow$ (owns) $V_3$ (method) $\rightarrow$ (call) $V_{\text{target}}$ This path is more useful because $V_3$ (method) can provide hints on how to call $V_{\text{target}}$, whereas the previous path included $V_2$ (script), which is a longer context and less useful for understanding how to use $V_{\text{target}}$.
\noindent In the training set of RepoBench-R, for each sample, we have a gold snippet and an in-file context. We first use kNN search to find the $k$ most similar context nodes to the in-file context. Then, we perform an exhaustive search from these identified nodes to collect all possible paths leading to the gold snippet. For example, in sample 1, we might collect paths of types (3, 4, 1) and (1, 2, 3), while in sample 2, we might find paths of types (3, 4, 1) and (2, 4). The most frequent path type in the training set is (3, 4, 1), which appears twice in this example.

\subsection{Context Types in Retrieved Node Analysis}
\label{sec:appendix-ct}
We follow definitions of different contex types in \citep{shrivastava2022repositorylevel} :
\begin{itemize}
    \item Parent Class (ParCls): code snippet taken from the parent class of the class that is having the target prediction line inside.
    \item Child Class (ChiCls): code snippet taken from the child class of the class that is having the target prediction line inside.
    \item Sibling (Sib): any code snippet in the files that are in the same directory as the current file (the file contains current target prediction line)
    \item Similar Name (SimN): take code snippet from files, functions, classes (objects) that have a similar name as the completing function, class or file. Similar names are determined by splitting the file name based on underscore or camel- case formatting and then matching parts of the filename. If one or more parts matches, two objects are considered to have similar names.
    \item Import Sibling (ImpSib): take code from the import objects (classes, functions) in the sibling files.
    \item Import Similar Name (ImpSimN): take code snippet from the import files used in the similar name files.
    \item Import Parent Class (ImpParCls): any code snippet from the import objects used in the parent class files. This implies the case when the child class is likely to re-use imported objects of its parent class.
\end{itemize}
\subsection{RepoBench Benchmark Statistics}
\begin{table*}[t]
\small
\centering
\begin{tabular}{@{}cccccccc@{}}
\toprule
Lang. & Task & Subset & XF-F & XF-R & IF & Mean Candidates & Mean Tokens \\
\midrule
\multirow{6}{*}{Python} & \multirow{2}{*}{RepoBench-R} & \hspace{6pt}Easy & 12,000 & 6,000 & - & 6.7 & - \\
& & \hspace{6pt}Hard & 12,000 & 6,000 & - & 17.8 & - \\ \cmidrule{2-8}
& \multirow{2}{*}{RepoBench-C} & \hspace{6pt}2k& 12,000 & 5,000 & 7,000 & - & 1,035 \\
& & \hspace{6pt}8k & 18,000 & 7,500 & 10,500 & - & 3,967 \\ \cmidrule{2-8}
& RepoBench-P & & 10,867 & 4,652 & 6,399 & 24 & 44,028 \\
\bottomrule
\\
\end{tabular}
\begin{tabular}{cccccc}
\toprule
Language & Task & XF-F & XF-R & IF \\
\midrule
\multirow{2}{*}{Python} & Code Retrieval & 175,199 & 86,180 & - \\
& Code Completion & 349,023 & 179,137 & 214,825 \\
\bottomrule
\end{tabular}
\caption{\textit{(Top)} Test data overview for RepoBench dataset for Python in 3 different tasks; \textit{(Bottom)} Training data for RepoBench for Python}
\label{tab:test_data_overview}

\vspace{-8pt}
\end{table*}

\subsection{Rerank as Link Prediction}
\label{rerank}
After extracting the subgraph using the k-nearest neighbor (kNN) search and expansion strategy, it is crucial to incorporate the query node, representing the in-file context, into the subgraph. To enable meaningful message passing between the query node and the extracted subgraph nodes, we need to establish edges connecting the query node to the subgraph. Fortuitously, in most cases, the query node (in-file context) already possesses certain connections to other nodes within the subgraph. For instance, the in-file context might invoke some functions that are represented as nodes in the subgraph for the task of next line prediction. By including the query node and its existing connections to the subgraph, the message passing network can effectively leverage the in-file context information during the link prediction process, consequently leading to more accurate and relevant code completion suggestions.
This integration of the query node into the subgraph allows for a comprehensive understanding of the code context, facilitating the message passing network to make informed predictions. By considering both the subgraph extracted from the repository and the in-file context, the model can capture the relevant relationships and dependencies, resulting in code completion suggestions that are tailored to the specific code context at hand.

\subsection{Samples}
\label{sec:appendix-samples}
In this section, we present some samples in which the similarity-based retrieval method failed at and why, we use the UniXCoder as the main encoder. These samples are collected in RepoBench-R Cross-File First subset.

\begin{figure}[t]
	
	\centering
    \includegraphics[width=\linewidth]{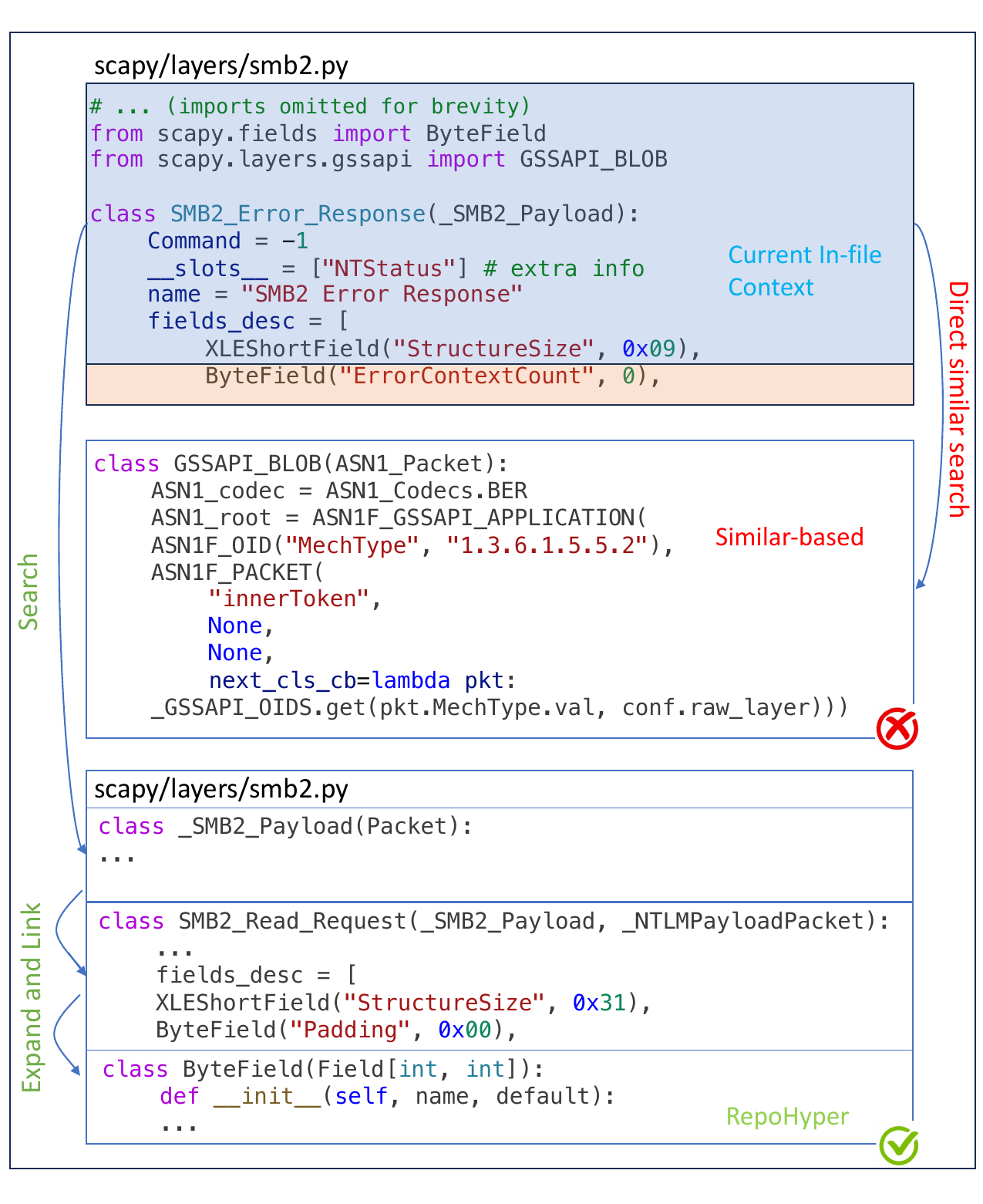}
    \caption{Sample ID 1430 in repository secdev/scapy}
    \label{fig:enter-label}
\end{figure}
\end{document}

%% file: math_commands.tex
\usepackage{amsmath,amsfonts,bm}

\def\eqref#1{equation~\ref{#1}}

\def\1{\bm{1}}

\DeclareMathAlphabet{\mathsfit}{\encodingdefault}{\sfdefault}{m}{sl}
\SetMathAlphabet{\mathsfit}{bold}{\encodingdefault}{\sfdefault}{bx}{n}



%% file: latex/introduction.tex
\section{Introduction}
\begin{figure}[t]
	
	\centering
	\includegraphics[width=\linewidth]{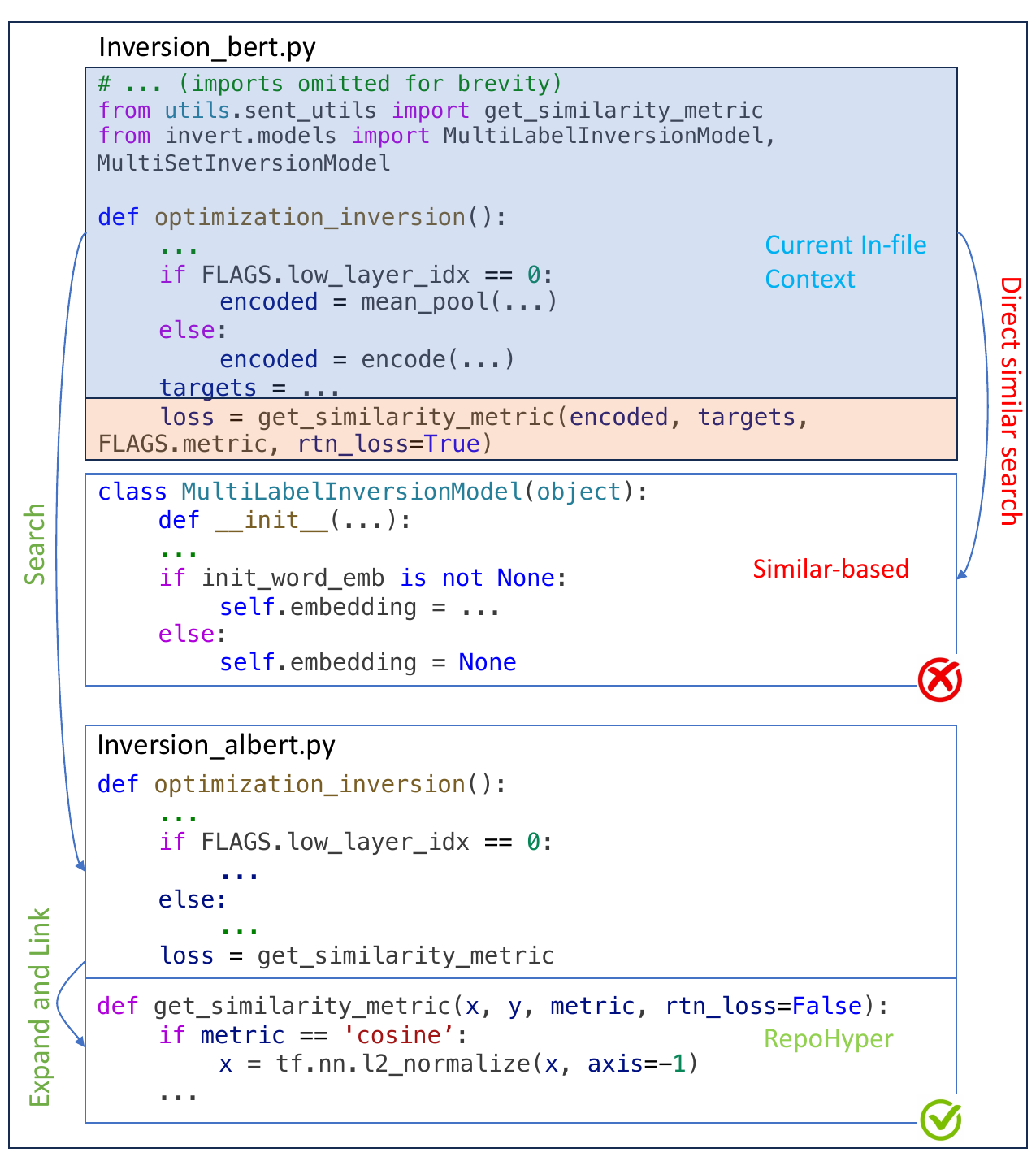}
	\caption{Illustration of graph-based semantic search versus similarity-based search. The orange block indicates the ground-truth line that needs to complete to call the function \textit{get\_similarity\_metric}. Similarity-based methods mistakenly focus on \textit{MultiLabelInversionModel} class due to its similarity in form with current in-file context, leading to incorrect completions. Conversely, \tool successfully identifies the correct context via first identify the most similar code snippet in the codebase then expand and link.}
	\label{fig:motiv}
\end{figure}

The advent of AI-assisted code completion tools, such as GitHub Copilot, marks a significant milestone in software development. These tools, while adept at interpreting the immediate context of the code being written, often do not fully exploit the broader context available within the entire code repository. This oversight results in suggestions that might not be optimally aligned with the project's architecture or intended functionality, as these tools tend to overlook the rich information embedded in related files, class hierarchies, dependencies, and more.

To overcome these shortcomings, a direct but complex solution involves enhancing the context length of language models by applying efficient attention techniques~\cite{dao2022flashattention,dao2023flashattention,press2021train,chen2023extending}. Nonetheless, increasing the context length significantly raises costs and is not feasible indefinitely, especially with respect to the voluminous number of files in a given repository. Hence, there is a crucial need for more refined strategies that accurately identify relevant contexts rather than indiscriminately analyzing every file within a repository. In response to this challenge, the concept of repository-level code completion has gained traction. It aims to incorporate the full context of a project, including inter-file relationships, imported modules, and the overarching project structure~\cite{liu2023repocoder, liao2023context, shrivastava2023repofusion, agrawal2023guiding, shrivastava2023repository}. These methodologies generally employ a similarity-based approach to retrieve contexts for completing a given code snippet, drawing either from raw source code or a pre-constructed database with essential metadata. However, this strategy exhibits significant limitations. It often fails to consider that diverse contexts within the current repository, not necessarily involving similar code, can provide valuable insights for code completion. This includes the intricate network of dependencies, shared utility functions, inter-module method calls, class hierarchies, inter-class dependencies, and encapsulation patterns—all of which are fundamental to program semantics.

To address these shortcomings, we introduce {\tool}, a novel repository-level code completion approach that considers global, repository-level contexts including not only similarity code but also program-semantic related contextual information within the current repository. Specifically, {\tool} incorporates three following components: (1)~\textbf{Repo-level Semantic Graph (RSG)}, which is a graph-structure designed to encapsulate the core elements of a repository's global context and their dependencies pertaining to code completion, serve as a reliable knowledge source to retrieve accurate contexts instead of raw codebase; (2)~\textbf{Expand and Refine} retrieval method which consists of two steps: (2.1)~\textbf{Search-then-Expand Strategies},which broaden the exploration of contexts by identifying semantically similar ones and then expanding the search to include contexts semantically linked, utilizing the RSG for intelligent navigation and (2.2)~\textbf{Link Predictor} which is a mechanism that refines the broad set of contexts obtained from the \textit{Search-then-Expand Strategies} and prioritizes them in a smaller, highly relevant subset for code completion. This is accomplished by formulating the re-ranking problem as a link prediction within the RSG, thereby reducing distractions for the LLM.

We conduct comprehensive evaluations of \tool on both \textbf{context retrieval (CR)} and \textbf{end-to-end code completion tasks (EECC)} using the RepoBench benchmark~\cite{liu2023repobench}, demonstrating significant improvements over existing state-of-the-art methods. In CR, \tool outperforms similarity-based approaches by an average of 49\% in retrieval accuracy, utilizing the same encoder~\cite{wang-etal-2023-codet5,guo2022unixcoder}. For EECC, our method surpasses RepoCoder and other RepoBench baselines, achieving an improvement of +4.1 in Exact Match (EM) and +5.6 in CodeBLEU scores.

To summarize, our main contributions are:
~\begin{enumerate}[leftmargin=*]
	\item We introduce \textbf{\tool}, a novel framework featuring three novel modules designed to address the multifaceted challenges of repository-level end-to-end code completion. 
	\item We develop the \textbf{RSG}, a novel graph representation that captures the global context of a repository, expanding to include non-similar but yet relevant contexts for repo-level code completion.  This innovation significantly improves the accuracy and relevance of context retrieval, surpassing conventional methods.
	\item We implement an \textbf{Expand and Refine} retrieval method via \textbf{Search-then-Expand Strategies} and \textbf{Link Prediction algorithm} within the RSG, optimizing the retrieval of the most relevant and program-semantic related contexts.
	\item We perform extensive evaluation of \tool in both repository-level code retrieval and code completion tasks demonstrates a significant improvement over the state-of-the-art approaches. Through a series of analytical and ablation studies, we confirm the vital role of each component of \tool.
\end{enumerate}

%% file: latex/relatedwork.tex
\section{Related Work}
\label{sec:relatedwork}
\subsection{Code LLMs for code generation \& understanding}

Recent research has introduced a plethora of Large Language Models (LLMs) tailored for code-related tasks~\cite{bui2023codetf, chowdhery2023palm, codex, austin2021program, hendrycks2021measuring, nijkamp2023codegen, nijkamp2023codegen2, zheng2023codegeex, wang-etal-2023-codet5, bui2018hierarchical, jayasundara2019treecaps, bui2023codetf, guo2024deepseekcoder, li2023starcoder, roziere2023code}, aiming to enhance code understanding and generation. These models are categorized into closed-source and open-source variants. Initially, closed-source models like Codex~\cite{codex}, Code-Davinci~\cite{codex}, and PaLM-Coder~\cite{chowdhery2023palm} demonstrated exceptional performance on well-known code completion benchmarks, including HumanEval~\cite{codex}, MBPP~\cite{austin2021program}, and APPS~\cite{hendrycks2021measuring}. Subsequently, the emergence of open-source models such as the CodeGen series~\cite{nijkamp2023codegen, nijkamp2023codegen2}, CodeT5~\cite{wang2021codet5}, CodeT5+~\cite{wang-etal-2023-codet5}, CodeGeeX~\cite{zheng2023codegeex}, StarCoder~\cite{li2023starcoder}, Wizard Coder~\cite{luo2023wizardcoder}, CodeLlama~\cite{roziere2023code}, and DeepSeek-Coder~\cite{guo2024deepseekcoder} began to rival the closed-source models in terms of benchmark performance. Despite their purported efficacy across a broad spectrum of code intelligence tasks, code generation and completion emerge as their most notable and widely utilized applications.


\subsection{Repository-level Code Completion}
Repository-level code completion has seen notable advancements through works like RLPG~\cite{RLGP}, CoCoMIC~\cite{ding2022cocomic}, RepoCoder~\cite{liu2023repocoder}, CodePlan~\cite{bairi2023codeplan} and A3-Codegen~\cite{liao2023context}. These studies highlight the critical role of leveraging both within-file and cross-file contexts to improve code completion accuracy. Conversely, RepoFusion~\cite{shrivastava2023repofusion}, RepoPrompts~\cite{shrivastava2023repository}, and MGD~\cite{agrawal2023guiding} propose methodologies for effectively integrating these contexts, assuming their availability from external sources. RepoBench~\cite{liu2023repobench} and CrossCodeEval~\cite{ding2023crosscodeeval} emphasize the need for end-to-end benchmarks designed to evaluate code completion systems within the broader, repository-level contexts. Furthermore, CodeAgent~\cite{zhang2024codeagent} incorporates documentation, contexts, runtime environments, and a pipeline for interacting with repositories through multi-agent systems.

%% file: latex/methodology.tex
\section{Methodology}
\label{sec:methodology}

\begin{figure*}[t]
    \centering
    \includegraphics[scale=0.124]{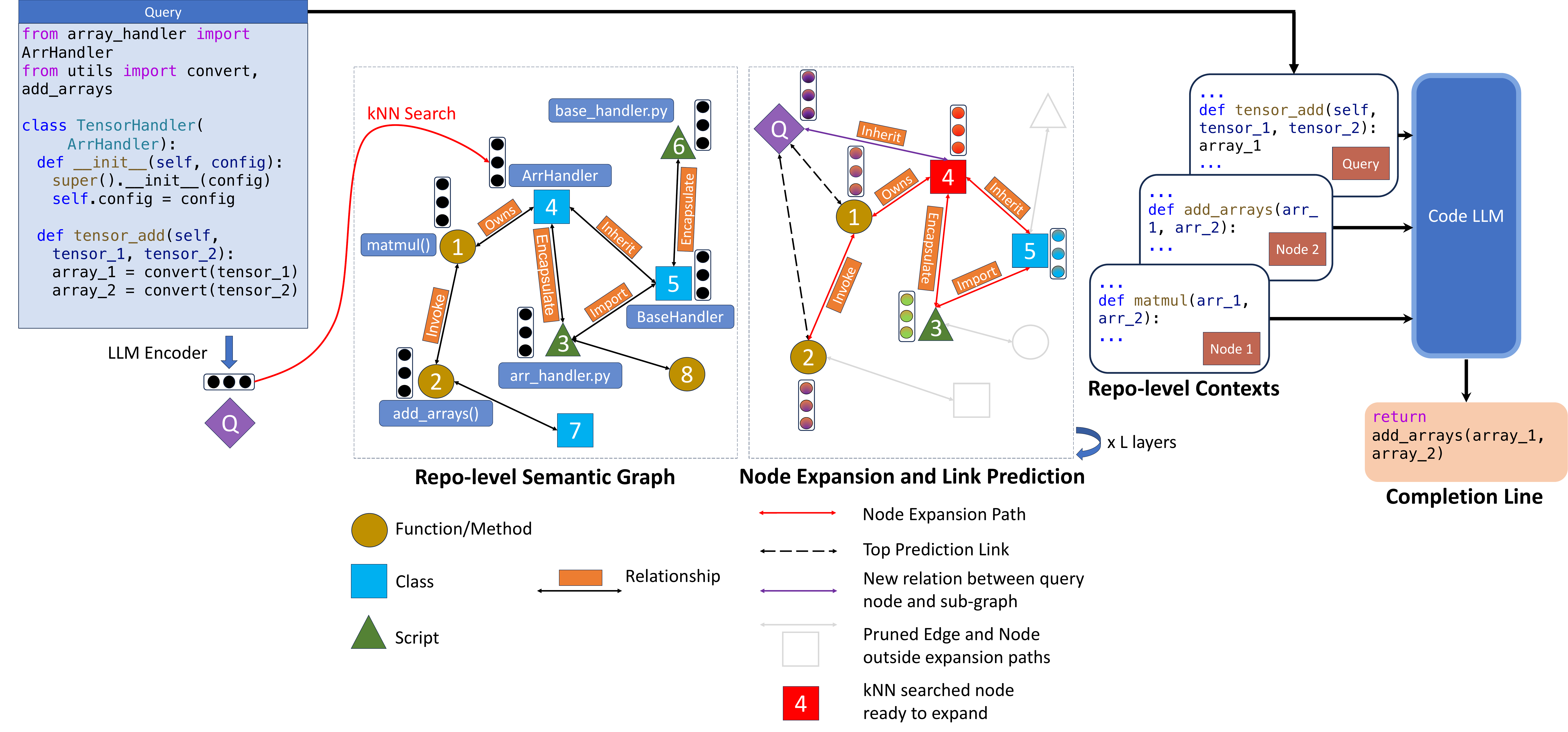}
    \vspace{-15pt}
    \caption{Overall Architecture of RepoHyper. Here we use $K=1$.}
    \label{fig:arch}
\end{figure*}

Figure~\ref{fig:arch} illustrates the overall architecture of our approach, {\tool}. 
Given an existing, incomplete code snippet $Q$, we first encode it into a semantic vector using an encoder function. Our objective is to retrieve relevant semantic contexts $T$ from the repository $R$. Consequently, these retrieved contexts are subsequently utilized by a Large Language Model (LLM) to generate the final code prediction:
\begin{equation*}
\small
C = \text{LLM}(Q, T).
\end{equation*}

We present a {\em Repo-level Semantic Graph} (RSG) for global context representation (Section~\ref{sec:graphrep}) and a {\em Expand and Refine} retrieval algorithm to re-rank and select relevant snippets from RSG. This includes two key steps: \textbf{Search-then-Expand} which tries to find the most similar and program-semantic related contexts, and \textbf{Re-ranking as Link Prediction} which aims to refine the contexts set found by the prior step. (Section~\ref{sec:search-expand} and Section~\ref{sec:link-prediction}). 


\subsection{Repo-level Semantic Graph (RSG): Representation for Global Contexts}
\label{sec:graphrep}


We denotes Repo-level Semantic Graph (RSG) as $ \mathcal{G} = (V,E)$, where $V$ denotes a set of nodes and $E$ denotes a set of relations, aims to capture the fundamental units of a project's global context and the intricate relationships among them. We consider (1) {\em function/method} and (2) {\em class} to be fundamental units due to their crucial role in program structure. Each node contains the name, parameters, and body of the corresponding function/method. This allows for precise context access and clear separation of function calls and class-method relationships, which is crucial for repository-level code completion \cite{shrivastava2022repositorylevel}. This also ensures precise context segmentation, eliminating the need for manual chunking and size tuning. After extracting functions and classes, the remaining file content, such as import statements and non-functional code, is encapsulated in a (3) {\em Script} node.

The nodes in an RSG are interconnected based on their types and relations, which are categorized as follows:
(1) {\em Import Relations}: These relations (Imports and Imported By) exist between script nodes and the imported modules identified from the script's import statements. This excludes external modules not within the project's scope; (2) {\em Invoke Relations}: These relations (Caller and Callee) exist between functions (or methods) when one node invokes another; (3) {\em Ownership Relations}: These relations (Owns and Owned By) exist between methods and the classes that contain them; (4) {\em Encapsulate Relations}: These relations (Enclose and Enclosed By) exist between script nodes and other nodes that have code snippets contained within the file represented by the script node; and (5) {\em Class Hierarchy Relations}: These relations (Inherits and Inherited By) exist between classes.


While our implementation is for Python, RSG can be adapted for using in other high-level programming languages, e.g., Java or C++. This adaptation involves utilizing different program elements as context units, and their relations in an RSG.

\subsection{Search-then-Expand Strategies}
\label{sec:search-expand}

Our methodology employs a search-then-expand approach to identify the most suitable file for the decoding process within a repository context. We aim to broaden the search space beyond similarity-based candidates to encompass more relevant files, as suggested by \citet{shrivastava2022repositorylevel, liu2023repocoder, liu2023repobench}. These studies indicate that files with similar imports, names, or code snippets are typically the correct contexts for retrieval, and that semantic search using k-Nearest Neighbor (kNN) with encoders like UniXCoder \citep{guo2022unixcoder} or CodeT5+ \citep{wang-etal-2023-codet5} is effective. Additionally, they highlight the importance of structured context sources such as \textit{Sibling} files or \textit{Import of Parent Class} in providing relevant contexts.

Based on these insights, {\tool} initially performs a kNN search with a small $K$ to identify a set of anchor nodes in the RSG. These nodes are then expanded using strategy $\mathcal{F}$:
\[
    A_{\text{exp}} = \mathcal{F}(A), \ \ A = \{V_i | i \in \textup{kNN} (\mathcal{G}, Z_Q)\}
\]
Here, $\text{kNN}(\mathcal{G}, Z_Q)$ represents the kNN search for the $K$ nodes most similar to the query vector $Z_Q$ in the graph $\mathcal{G}$. We experiment different strategies to find the most optimal nodes in the graph for decoding process. Our two proposed strategies are:

(1) \textit{Exhausted Search}: Beginning from a node $V_j \in A$, we utilize a straightforward Breadth First Search algorithm (BFS) with a maximum depth of $D$. Theoretically, $D$ should reach 4 to encompass the complete relationship between two contexts for repo-level code completion. However, in practice, setting $D \geq 3$ may result in the BFS covering nearly 50\% of the graph. Hence, we introduce another parameter alongside $D$ to constrain the number of BFS expanded nodes: the maximum number of nodes per BFS denoted as~$M$.

(2) \textit{Pattern Search}: We found that exhausted search on all the directions and paths can include too many irrelevant contexts to the query. Thus, we conduct kNN search for all the queries in the training set, then expand using exhausted strategy from kNN searched node $V_j$ to target node $V_{\text{target}}$, then collect the most frequent type paths into path set $\mathcal{P}$, then add them as filters for later BFS. This is called as pattern search since it will eliminates non-frequent paths during exhausted exploration saving the walking nodes.
$$
    A_{\text{exp}} = \{V_i | V_i \in \mathcal{F}_{\text{exh}}(A), \text{PATH}(V_j, V_i) \in \mathcal{P} \}
$$
where $\text{PATH}$ denotes the type of BFS exploration path from kNN searched node $V_j \in A$ to any walking node $V_i$. For example, one might prioritize the exploration from a class node to its script node then to imported function to draw a possible invoke relationship for code completion (class-script-function path type), but one is unlikely to explore chain of method calls (method-method-method path type). More details are in Appendix \ref{appendix:ps}.

\subsection{Re-ranking as Link Prediction}
\label{sec:link-prediction}


To manage the increased number of contexts from our Search-then-Expand strategy and reduce noise, we refine the context selection process for the decoder. Instead of using all retrieved contexts, we only consider the top-$N_2$ (where $N_2 < N_1$) most relevant ones. Initially, we tried ranking contexts by their embedding distances to the query, but this approach underperformed in our evaluation.

To improve relevance ranking, we treat it as a link prediction problem on a RSG. We embed the query as a node in the RSG and use a message passing network, along with a link prediction head, to score the connections between the query and other nodes in the graph. The final embeddings after going to message passing network $f$ are used to calculate linking scores, which determine the relevance of each context to the query.

\begin{equation*}
\small
    Z_i^{(L)} = f(Z_i^{(0)}, \mathcal{G}\oplus \mathcal{Q})
\end{equation*}
$Z_i^{(0)}$ is initial embedding taken from encoder for $i^{\text{th}}$ node in graph, $Z_i^{(L)}$ is the last layer embedding after going to message passing network~$f$. $\mathcal{G}_1= \mathcal{G}\oplus \mathcal{Q}$ is concatenation of query $\mathcal{Q}$ node has query $Q$ as the raw source code of the node and $Z_Q$ as initial embedding to $\mathcal{G}$, by adding this into set of nodes $V$ and add new relations to the query. For example, if the current code already have invoke relations with some function nodes in graph, we add these edges to the current node avoiding duplicate prediction.
\begin{equation}
\small
    s_i=W^T \mathrm{concat}(Z_i^{\left(L\right)}, Z_Q^{\left(L\right)}) \ \ \forall i \in  \{i | V_i \in A_{\text{exp}}\}
\end{equation}

, where $W$ is a trainable model parameter, $s_i$ is the linking score between query node and all other nodes inside the RSG. In practice, we focus on nodes that are imported into the file being predicted, as suggested in RepoBench \cite{liu2023repobench}. The training loss of node link prediction for each query is computed as
\begin{equation}
\small
\mathcal{L} = -\frac{1}{N_1} \sum_{i=1}^{N_1} y_i \log \hat{y}_i \textup{ where } \hat{y}_i = \frac{1}{1+e^{-s_i}} 
\end{equation}
More details about incorporating query node into inference process can be found in Appendix. \ref{rerank}.
In all experiments, we employ GraphSAGE \cite{hamilton2017inductive} with $L$ layers as the graph neural network (GNN) model to update the representation for each node based on the passage graph. The $l$-th layer of the GNN model updates the embedding of node $i$ as follows:
\begin{equation}
\small
Z_i^{(l)}=h\left(Z_i^{(l-1)},\left\{Z_j^{(l-1)}\right\}_{(i, j) \in \mathcal{G}_1}\right)
\end{equation}
where $h$ is usually a non-linear learnable function which aggregates the embeddings of the node itself and its neighbor nodes. 
After re-ranking with linking scores, the final top-$N_2\left(N_2<N_1\right)$ contexts are sent for decoding. Suppose their indices are $\left\{g_1, g_2, \cdots, g_{N_2}\right\}$, the decoding process for final prediction $\mathbf{C}$ is:
\begin{equation}
\small
\mathbf{C}=\text { LLM }\left(Q, \left[\mathbf{P}_{g_1} ; \mathbf{P}_{g_2} ; \cdots ; \mathbf{P}_{g_{N_2}}\right] \right)
\end{equation}
where $\mathbf{P}_{i}$ is code representation of node $V_{i} \in V$ 

To train such a network to re-rank inside a repository, we solely optimize loss function defined in (1) on dataset $R = \left \{ \left(\mathcal{G}^i, y^{i}_{\text{optimal}}, Q^i \right) \right \}$ with $\mathcal{G}^i$ is a RGS for a $i^\text{th}$ repository and $y^{i}_{\text{optimal}}$ is the optimal context node corresponding to query $Q^i$.

%% file: latex/experiment-setup.tex
\section{Empirical Evaluation}

\input{latex/tables/retrieval}

\label{sec:exp-setup}
\subsection{Tasks \& Datasets}
We choose RepoBench~\cite{liu2023repobench}as the main dataset for our  evaluation pipeline due to its large scale and comprehensive, making it an ideal candidate for assessing repository-level code completion from multiple perspectives. RepoBench~consists of three distinct subsets, each designed to evaluate different aspects of repo-level code completion. Each subset contains up to 12000 samples for evaluation~\footnote{Statistic of RepoBench can be found in Appendix Section.}.
\begin{enumerate}[leftmargin=*]
	\item \textbf{RepoBench-R} focuses on evaluating the retrieval of relevant code snippets, crucial for accurate code prediction. It assesses the model's ability to sift through extensive repository data to identify useful snippets for code completion, termed as the \textbf{Context Retrieval} task.
	\item \textbf{RepoBench-C} is designed to predict the next line of code using provided in-file and cross-file contexts, testing a model's ability to predict precise code completion from available contexts.
	\item \textbf{RepoBench-P} combines the challenges of RepoBench-R and C, testing a model's pipeline from snippet retrieval to code prediction, reflecting real-world auto-completion (denoted as the \textbf{End-to-End Code Completion} task).
\end{enumerate}

Our work focuses on the creation of repository-level code graphs and retrieval strategies, primarily utilizes RepoBench-R and RepoBench-P. These subsets align with our goals to enhance context retrieval and code completion accuracy, directly showcasing our contributions.

Within these benchmarks, there are two settings \cite{liu2023repocoder} to thoroughly assess performance: \textbf{Cross-File-First} (XF-F) challenges a model to predict the first occurrence of a cross-file line, requiring adept handling of long-range contexts; \textbf{Cross-File-Random} (XF-R) masks a random cross-file line where prior usage might offer clues for prediction. These settings enable a robust evaluation on these code completion scenarios.

\subsection{Baselines \& Metrics}
\subsubsection{Context Retrieval on RepoBench-R}
We follow ~\citet{liu2023repobench} to define 4 baselines: (1) \textbf{Random Retrieval}, where code snippets are chosen randomly, providing a basic comparison level. This process is repeated 100 times to average the results for consistency. (2) \textbf{Lexical Retrieval}, which uses simple text comparison techniques like Jaccard Similarity and Edit Distance to find relevant snippets based on the code tokens. (3) \textbf{Similarity-based Retrieval}, employing encoder models like CodeBERT, UnixCoder and OpenAI Text Embedding Models to generate code embeddings and Cosine Similarity to measure the semantic similarity between the cropped code and the candidate snippets. In \tool setting, we employ our pipeline, which leverages different encoder models (UniXCoder~\cite{guo2022unixcoder}, CodeT5+-770M~\cite{wang-etal-2023-codet5}) to encode the query, construct semantic graph, node expansion, and link prediction to retrieve the contexts.

\paragraph{Evaluation Metrics} We also follow ~\cite{liu2023repobench} use to  Accuracy@k (acc@k) metric to evaluate performance on this task. For the easy subset of tasks, we assess performance using acc@1 and acc@3, while for the more challenging subset, we evaluate using acc@1, acc@3, and acc@5.

\subsubsection{Code Completion on RepoBench-P}
This end-to-end code completion task requires two components: Context Retrieval and Code Completion. Given our research's emphasis on developing novel context retrieval strategies, we examine how these strategies perform under various settings, using consistent code completion models for comparison. Specifically, we utilized GPT-3.5-Turbo and DeepSeek-Coder-33B~\cite{guo2024deepseekcoder} as our the LLM for code completion in our workflow.

We follow ~\cite{liu2023repobench} to define baselines according to different settings on the contexts: (1) \textbf{Gold-Only}, which uses only the `gold snippet' for cross-file completions and leaves in-file completion contexts empty, testing a model's efficacy with minimal context. (2) \textbf{In-File-Only}, which uses maximum 30 lines up from prediction line in same file, indicates lower-bound performance without repo-level contexts. (3) \textbf{RepoBench}, which uses similarity-based search, UniXCoder, to retrieve top contexts, then prompts LLM for code completion. 

We also assess snippet ranking strategies, \textbf{H2L} (High-to-Low) and \textbf{L2H} (Low-to-High), to see how the order of context relevance affects code completion performance. We also include \textbf{RepoCoder}~\cite{liu2023repocoder}, which is a method on the repo-level code completion task as another baseline. This method is run iteratively with Jaccard retrieval method and maximum iterations of~4.

\paragraph{Evaluation Metrics} We use Exact
Match (EM) and CodeBLEU~\cite{ren2020codebleu} to measure next-line completion accuracy as in~\cite{liu2023repobench}.


\subsection{Implementation Details}
To construct the Repo-level Semantic Graph (RSG) for a repository, we first parse functions, methods, and classes using \textit{tree-sitter}\footnote{\url{https://github.com/tree-sitter/tree-sitter}}, a tool for generating abstract syntax trees (AST). We then extract code entities from the AST and integrate them into a semantic graph\footnote{See Appendix \ref{sec:appendix-rsg} for RSG construction details.}. In our methodology, we adopt \textit{Pattern Search} for expansion with parameters set to a maximum depth $D=4$, $M=1000$, and $K=3$. The number of prompt contexts for LLM completion, $N_2$, is dynamically selected to maximize token count within LLMs' context length limits. The Link Predictor is trained on training subset of RepoBench-R's gold context labels, with a text matching algorithm based on Jaccard distance used to align labels with RSG nodes. We use a homogeneous GraphSAGE network $f$ with $L=3$ GNN layers, optimized with Adam at a learning rate of $0.01$ for 10 epochs, noting homogeneous networks performed adequately for RSG.  Link Predictor is trained on 2 A100 GPUs in 6 hours.

%% file: latex/tables/retrieval.tex
\begin{table*}[t]
\setlength{\tabcolsep}{4.5pt}
\centering
\small
\begin{tabular}{llccccrccrcc}
\toprule
   \multirow{3}{*}{Retrieval} & \multirow{3}{*}{Model} & \multicolumn{4}{c}{Easy} & \multicolumn{6}{c}{Hard} \\
\cmidrule(lr){3-6} \cmidrule(lr){7-12} 
    & & \multicolumn{2}{c}{XF-F} & \multicolumn{2}{c}{XF-R} & \multicolumn{3}{c}{XF-F} & \multicolumn{3}{c}{XF-R} \\
\cmidrule(lr){3-4} \cmidrule(lr){5-6} \cmidrule(lr){7-9} \cmidrule(lr){10-12}
   &  & acc@1 & acc@3 & acc@1 & acc@3 & acc@1 & acc@3 & acc@5 & acc@1 & acc@3 & acc@5 \\
\midrule
 Random & & 15.68 & 47.01 & 15.61 & 46.87 & 6.44 & 19.28 & 32.09& 6.42 & 19.36 & 32.19 \\
\cmidrule(lr){1-12}  
 \multirow{2}{*}{Lexical} & Jaccard & 20.82 & 53.27 & 24.28 & 54.72 & 10.01 & 25.88 & 39.88& 11.38 & 26.02 & 40.28 \\
& Edit & 17.91 & 50.61 & 20.25 & 51.73 & 7.68 & 21.62 & 36.14& 8.13 & 22.08 & 37.18 \\
\cmidrule(lr){1-12}  
 \multirow{2}{*}{Similarity-based} & CodeBERT & 16.47 & 48.23 & 17.87 & 48.35 & 6.56 & 19.97 & 33.34& 7.03 & 19.73 & 32.47 \\
  & UniXcoder & 25.94 & 59.69 & 29.40 & 61.88 & 17.70 & 39.02 & 53.54 & 20.05 & 41.02 & 54.92 \\
 & CodeT5+ & 18.29 & 53.31 & 19.61 & 53.05 & 9.51 & 25.24 & 37.81& 13.21 & 28.15 & 38.76 \\

 & OpenAI & 28.14 & 64.24 & 31.15 & 65.29 & 18.23 & 42.01 & 60.39 & 23.78 & 45.67 & 58.53 \\
 
\cmidrule(lr){1-12}  
 \multirow{2}{*}{{\tool}} & UniXcoder & \textbf{32.56} & \textbf{68.91} & \textbf{33.79} & \textbf{69.67} & 
 \textbf{25.81} & \textbf{49.51} & \textbf{61.19} &\textbf{27.12} & \textbf{51.12} & \textbf{63.23}  \\
 & CodeT5+ & 31.15 & 67.23 & 32.01 & 68.12 & 25.16 & 46.70 & 60.19 & 23.81 & 43.61 & 57.12 \\
\bottomrule
\end{tabular}
\caption{Results of \tool on RepoBench-R dataset. The studied encoder models include \texttt{codebert-base} for CodeBERT, \texttt{unixcoder-base} for UniXcoder, \texttt{codet5p} for using CodeT5+ 770M parameters. {\tool} has UniXcoder and CodeT5+ 770M as the base encoders, respectively. Numbers are shown in percentage (\%), with the best performance highlighted in bold. \texttt{text-embedding-large-3} for OpenAI.
}
\label{tab:repobench-r}
\end{table*}

%% file: latex/results-analysis.tex
\section{Evaluation Results}
\label{sec:results}

\subsection{Performance on Context Retrieval}


The results presented in Table \ref{tab:repobench-r} demonstrate the enhanced performance of {\tool} when compared to similarity-based approaches . By implementing graph-based semantic search strategies, {\tool} significantly outperforms the baseline methods, including those utilizing CodeBERT and UniXCoder, as well as our own tests with CodeT5+. Specifically, it achieves~high improvements, with UniXcoder and CodeT5+ showing relative increases of nearly 26\% and 72\%, respectively, across various subsets and task scenarios.

\subsection{Performance on Code Completion}

\input{latex/tables/code-completion}

Table~\ref{tab:p} shows that {\tool} emerges as the most effective strategy, consistently achieving the highest scores across both EM (Exact Match) and CodeBLEU metrics in both XF-F and XF-R settings, with notable scores such as 52.76\% EM and 61.49\% CodeBLEU with gpt-3.5-turbo, and 49.49\% EM and 58.48\% CodeBLEU with DeepSeek-Coder. Compared to the baseline strategies like Gold-Only, RepoBench-L2H/H2L, and RepoCoder, {\tool}'s strategies (both L2H and H2L) demonstrate superior performance. 
The comparison between L2H (Low-to-High relevance) and H2L (High-to-Low relevance) within {\tool} indicates that prioritizing snippets from low to high relevance (L2H) offers a significant advantage over the reverse, particularly  highlighting the high performance of {\tool}-L2H strategy.

\section{Ablation Study}
\subsection{Hyper-parameters Sensitivity and Effects}

\input{latex/tables/hyper_params}

We evaluated various graph-based semantic search strategies, including Exhausted, Pattern, and the baseline kNN search, each with different hyper-parameters. Table~\ref{tab:sensitivity-analysis} reveals that while Exhausted Search achieves high hit rates, Pattern Search offers a more efficient solution, reaching up to 73\% hit rates by exploring only 28\% of the nodes, compared to 36.7\% needed by Exhausted Search for similar outcomes. This efficiency highlights~Pattern Search's ability to identify relevant sub-graphs more efficiently, enhancing precision with less computational resources. Moreover, both Exhausted and Pattern Search strategies significantly outperform the kNN baseline in hit rates, while maintaining smaller, more focused sub-graphs for quicker inference and reduced noise. These results emphasize the importance of choosing the right search strategy and hyperparameter tuning to balance search thoroughness and prediction efficiency.

\subsection{Retrieved Nodes Analysis}
{\tool} aims to retrieve both semantically similar code contexts and program-semantic related contexts like \textit{Sibling} (Sib) or \textit{Import of Parent Class} (ImpParCls). In this study, we validated {\tool}'retrieval capability across various context types identified in RepoBench-R's \textit{Hard} subset, adhering to a classification scheme from prior research. We excluded rare context types due to their minimal representation in the dataset and focused our experiment on 100 instances for each context type under the XF-R setting, using the acc@5 metric for evaluation.

Results in Figure~\ref{fig:retrieval_performance} show that our model excels in retrieving contexts tied to global program semantics, e.g., Class Hierarchy, underscoring its effectiveness. However, it underperforms in retrieving contexts with similar names, likely due to the limited number of anchors used during the kNN search phase. Adjusting the number of anchors might improve retrieval of similar snippets but could impact other context types' retrieval efficiency.

\input{latex/figures/node_analysis_fig}

\subsection{Ablation Study}
\input{latex/tables/ablation_study}


Several meaningful observations can be drawn from Table \ref{table:ablation}: 
without Link Predictor, Pattern Search offers better accuracy than Exhausted Search. This is likely due to the fact that Exhausted Search has to include more nodes to obtain similar hits rate making the later re-ranking more difficult because of the noisy context nodes.

%% file: latex/tables/code-completion.tex
\begin{table}[h]
\centering
\small
\tabcolsep 2.5pt
\resizebox{\columnwidth}{!} {
\begin{tabular}{lllcccccc}
\toprule
  & \multirow{2}{*}{\textbf{CR Strategy}} & \multicolumn{2}{c}{\textbf{XF-F}}   & \multicolumn{2}{c}{\textbf{XF-R}}     \\ \cmidrule(lr){3-4} \cmidrule(lr){5-6}   & &   EM             & CodeBLEU             & EM                  & CodeBLEU           \\ \midrule
\multirow{9}{*}{\rotatebox[origin=c]{90}{\textbf{Gpt-3.5-turbo}}} 
& In-File-Only$^\ast$ & 26.35 & 33.14 & 36.31 & 44.01 \\
& Gold-Only$^\ast$ & 30.59 & 38.37 & 40.65 & 49.12 \\
\cmidrule{2-6}
& RepoBench-L2H & 37.51 & 46.19 & 49.3 & 57.20 \\
& RepoBench-H2L & 39.89 & 48.01 & 51.21 & 59.44 \\
\cmidrule{2-6}
& RepoCoder  & 48.73 & 57.62 & 59.55 & 67.42 \\

\cmidrule{2-6}

& {\tool}-L2H & \textbf{52.76} & \textbf{61.49} & \textbf{64.06} & \textbf{71.59}  \\

& {\tool}-H2L & 48.8 & 57.21 & 59.81 & 66.85 \\

\midrule

\multirow{9}{*}{\rotatebox[origin=c]{90}{\textbf{DeepSeek-Coder}}} 
& In-File-Only$^\ast$ & 25.46 & 32.39 & 34.23 & 42.13 \\
& Gold-Only$^\ast$ & 27.29 & 35.15 & 37.10 & 46.08 \\
\cmidrule{2-6}
& RepoBench-H2L & 34.02 & 43.08 & 45.96 & 53.62 \\
& RepoBench-L2H & 36.55 & 44.53 & 47.71 & 55.81 \\
\cmidrule{2-6}
& RepoCoder & 45.47 & 54.32 & 56.24 & 64.23 \\ 
\cmidrule{2-6}

& {\tool}-L2H & \textbf{49.49} & \textbf{58.48} & \textbf{59.03} & \textbf{66.98}  \\

 & {\tool}-H2L & 45.17 & 53.71 & 56.56 & 63.41 \\
\bottomrule
\end{tabular}
}
\caption{Comparison of various context retrieval strategies (CR Strategy) on the end-to-end code completion task on RepoBench-P for Python using GPT-3.5-turbo-16k and DeepSeek-Coder-33B.}.
\label{tab:p}
\vspace{-8pt}
\end{table}

%% file: latex/tables/hyper_params.tex
\begin{table}[h]
	\centering
	\small
	\resizebox{\columnwidth}{!}{
		\begin{tabular}{c|l|cc}
			\toprule
			\multirow{2}{*}{\textbf{Search Algorithm}} & \multirow{2}{*}{\textbf{Parameter Combination}} & \multicolumn{2}{c}{\textbf{Hit/Coverage}} \\
			\cmidrule{3-4} 
			& & \textbf{Hits} & \textbf{Coverage} \\
			\midrule
			\multirow{6}{*}{Exhausted Search} & D=4, M=1000, K=3 & 80\% & 40\% \\
			& D=4, M=200, K=4 & 78\% & 44\% \\
			& D=4, M=10000, K=1 & 72\% & 40.4\% \\
			& D=2, M=10000, K=1 & 34\% & 10\% \\
			& D=2, M=10000, K=2 & 47\% & 16\% \\
			& D=2, M=10000, K=8 & 73\% & 36\% \\
			\midrule
			\multirow{6}{*}{Pattern Search} & D=4, M=1000, K=3 & 73\% & 28\% \\
			& D=4, M=200, K=4 & 70\% & 29\% \\
			& D=4, M=10000, K=1 & 65\% & 31\% \\
			& D=2, M=10000, K=1 & 62\% & 14\% \\
			& D=2, M=10000, K=2 & 51\% & 8\% \\
			& D=2, M=10000, K=8 & 68\% & 21\% \\
			\midrule
			kNN & D=M=1, K=0.35*|$\mathcal{G}$| & 53\% & 35\% \\
			\bottomrule
		\end{tabular}
	}
    \caption{Sensitivity Analysis. $D$ : maximum depth; $M$ : maximum number of nodes; $K$ : in $\mathrm{kNN}$ algorithm. During expansion, the search algorithm will expand $K$ nodes to Coverage \% of the number nodes of the repolevel semantic graph, then extract explored node into a sub-graph. There's a Hits \% probability that there's an optimal context in this sub-graph. In kNN baseline, we use $K$=$35 \%$ of size of the original graph.}
	\label{tab:sensitivity-analysis}
\end{table}

%% file: latex/figures/node_analysis_fig.tex
\begin{figure}
\centering
\begin{tikzpicture}[scale=0.75]
\begin{axis}[
    ybar,
    bar width=0.22cm,
    width=7.5cm,
    height=6.4cm,
    enlarge x limits=0.25,
    legend style={
        at={(0.5,0.95)},
        anchor=north,
        legend columns=-1,
        font=\fontsize{7.5}{0}\selectfont,
        /tikz/every even column/.append style={column sep=0.5cm}
    },
    ylabel={Retrieval Performance Acc@5},
    ylabel style={font=\fontsize{9}{0}\selectfont},
    xlabel={Context Types},
    xlabel style={xshift=-8pt, font=\fontsize{9}{0}\selectfont}, 
    symbolic x coords={SimN,ParCls,ChiCls,Sib,ImpSib,ImpSimN,ImpParCls},
    xtick=data,
    x tick label style={
    rotate=45,
    anchor=east,
    align=right,
    font=\fontsize{9}{0}\selectfont},
    ymin=0,ymax=1.0,
    grid style=dashed,
    ymajorgrids=true
]
\addplot[fill=blue] coordinates {
    (SimN,0.61)
    (ParCls,0.75)
    (ChiCls,0.73)
    (Sib,0.61)
    (ImpSib,0.41)
    (ImpSimN,0.56)
    (ImpParCls,0.48)
};
\addplot[fill=orange] coordinates {
    (SimN,0.71)
    (ParCls,0.61)
    (ChiCls,0.63)
    (Sib,0.46)
    (ImpSib,0.24)
    (ImpSimN,0.35)
    (ImpParCls,0.31)
};
\legend{RepoHyper, Similarity-based Search}
\end{axis}
\end{tikzpicture}
\vspace{-6pt}
\caption{Retrieval performance comparison between {\tool} and Similarity-based Semantic Search across different context types. We use kNN search within our RSG with UniXCoder encoder for encoding, this method is denoted as Similarity-based Semantic Search and {\tool} with same encoder. Please see Appendix \ref{sec:appendix-ct} for more details on Context Types.}
\label{fig:retrieval_performance}
\end{figure}

%% file: latex/tables/ablation_study.tex
\begin{table}[h]
		\normalsize
		\centering
        \small

		\setlength{\tabcolsep}{2mm}{
		    \begin{tabular}{lcc}
				\toprule
				Models & Easy & Hard \\
				\midrule
				(1) kNN & 60.15&40.74\\
				(2) \ \ w/ \text{Exhausted Search}&62.35&42.88\\
				(3) \ \ w/ \text{Pattern Search}&64.23&44.50\\
				(4) \ \ w/ E.S+Link Predictor &68.10&47.15\\
				(5) \ \ w/ P.S+Link Predictor &69.12&47.83\\
				(6) \ \ w/ P.S+Re-ranking   &67.43&44.48\\
				\bottomrule
				
		    \end{tabular}
          \caption{Ablation study in the code retrieval task, we use Repobench-R testset with two \textit{Easy} and \textit{Hard} subsets. Acc@3 is used as the main metric. Exhausted Search is denoted as E.S, Pattern Search is denoted as P.S. For a combination with kNN and expansion strategy, we simply re-rank inside extracted sub-graph using cosine similarity between query and nodes in sub-graph.}
		    \label{table:ablation}
		    } 

\end{table}

%% file: latex/conclusion.tex
\section{Conclusion}

In this paper, we introduced \tool, a novel framework aimed at enhancing repository-level code completion by addressing its complex challenges. \tool advances this domain through three key components: the Repo-level Semantic Graph (RSG), Search-then-Expand Strategies, and a Link Predictor, collectively improving the accuracy and relevance of code suggestions. With extensive evaluation, \tool show superiority in Repo-level Code Completion.

%% file: latex/limitations.tex
\section{Limitations}



This section outlines the limitations of our study, which we hope will serve as a catalyst for further research in this field:

Firstly, our Pattern Expansion strategy within the Repobench-R training set involves collecting the most frequent path types from kNN searched nodes to the nearest target nodes, which are then incorporated into the path type set $\mathcal{P}$ and used as filters for BFS. During this process, we manually select path types for exploration, which may not be optimal and could vary across different programming languages. A potential solution to this issue is the design of a learnable algorithm for navigating in RSG, as suggested by \cite{moon-etal-2019-opendialkg}.

Secondly, our experiments on repository-level tasks were conducted using only the RepoBench dataset. Although, RepoBench is a well-designed benchmark with a sufficiently large sample size to statistically validate the effectiveness of our approach, the generalizability of our findings would be strengthened by performing detailed analyses on additional repository-level code completion benchmarks, such as those presented in \cite{ding2023crosscodeeval, liu2023repocoder}. 

Lastly, our experiments rely on both public and proprietary large-scale CodeLLMs, which necessitate significant computational resources and contribute to carbon emissions, as highlighted by \cite{patterson2021carbon}. Moreover, the predictions generated by these models may not always align with user intentions, a concern that is further discussed in \cite{chen2021evaluating}. Addressing these issues is crucial for developing more environmentally sustainable and user-aligned CodeLLMs in the future.